\documentclass[aps,prl,reprint, floatfix, 10pt, showpacs,superscriptaddress]{revtex4-1}  

\newcommand{\m}{\mbox{ m}}
\newcommand{\cm}{\mbox{ cm}}

\newcommand{\Hz}{\mbox{ Hz}}
\newcommand{\kHz}{\mbox{ kHz}}
\newcommand{\MHz}{\mbox{ MHz}}
\newcommand{\GHz}{\mbox{ GHz}}
\newcommand{\nsec}{\mbox{ nsec}}

\usepackage{graphicx}
\usepackage{epstopdf}
\usepackage{bbold}
\usepackage{comment}
\usepackage{color}
\usepackage{soul}
\usepackage[abs]{overpic}
\usepackage{amsmath,amsfonts,amssymb}
\usepackage{lipsum}
\usepackage{hyperref}
\usepackage{setspace}
\usepackage{siunitx}
\usepackage{bm}
\usepackage[bottom]{footmisc}

\makeatletter
\AtBeginDocument{\let\LS@rot\@undefined}
\makeatother

\graphicspath{{Figures/}}
\begin{document}


%
\author{Ronen Chriki}
\affiliation{Department of Physics of Complex Systems, Weizmann Institute of Science, Rehovot 7610001, Israel }
\author{Simon Mahler}
\affiliation{Department of Physics of Complex Systems, Weizmann Institute of Science, Rehovot 7610001, Israel }
\affiliation{Univ. Paris Sud, Universit\'e Paris Saclay, 91405 Orsay, France}
\author{Chene Tradonsky}
\affiliation{Department of Physics of Complex Systems, Weizmann Institute of Science, Rehovot 7610001, Israel }
\author{Vishwa Pal}
\affiliation{Department of Physics of Complex Systems, Weizmann Institute of Science, Rehovot 7610001, Israel }
\author{Asher A. Friesem}
\affiliation{Department of Physics of Complex Systems, Weizmann Institute of Science, Rehovot 7610001, Israel }
\author{Nir Davidson}
\affiliation{Department of Physics of Complex Systems, Weizmann Institute of Science, Rehovot 7610001, Israel }

\begin{abstract}
Spatial coherence quantifies spatial field correlations over time, and is one of the fundamental properties of light. Here we investigate the spatial coherence of highly multimode lasers in the regime of short time scales. Counter intuitively, we show that in this regime, the temporal (longitudinal) modes play a crucial role in spatial coherence reduction. To evaluate the spatial coherence we measured the temporal dynamics of speckle fields generated by a highly multimode laser with over $10^5$ lasing spatial (transverse) modes, and examined the dependence of speckle contrast on the exposure time of the detecting device. We show that in the regime of short time scale, the spatial and temporal modes interact to form spatio-temporal supermodes, such that the spatial degrees of freedom are encoded onto the temporal modes. As a result, the speckle contrast depends on the number of temporal modes, and the degree of spatial coherence is reduced and the speckle contrast is suppressed. In the regime of long times scale, the supermodes are no longer a valid representation of the laser modal structure. Consequently, the spatial coherence is independent of the temporal modes, and the classical result, where the speckle contrast is suppressed as the number of spatial modes, is obtained. Due to this new spatio-temporal mechanism, highly multimode lasers can be used for speckle suppression in high-speed full-field imaging applications, as we demonstrate here for imaging of a fast moving object.
\end{abstract}

\title[]{Spatio-Temporal Supermodes:\\Rapid Reduction of Spatial Coherence in Highly Multimode Lasers}
\maketitle


\section{I. Introduction}
Spatial coherence is a fundamental property of laser light, whereby the divergence of collimated light of high coherence is small and the light can be focused to small diffraction limited spots. However, high spatial coherence can cause deleterious effects, such as speckle noise. Speckles are erratic intensity variations formed by interference between waves, whose phase differences are random. In optical imaging systems, such uncontrolled scattering causes random interference, resulting in noisy speckle patterns that reduce signal to noise ratio (SNR) and corrupt the output image~\cite{Goodman2005a, Goodman2015}.

\begin{figure*}
    \includegraphics{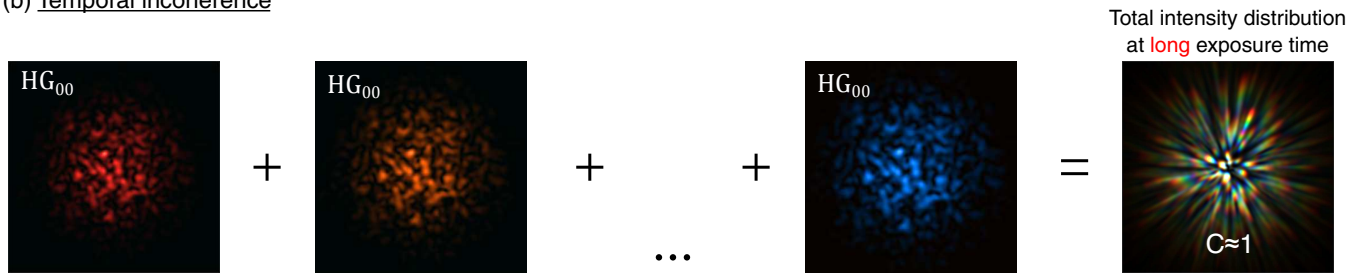}
	\caption{Comparison of speckle suppression by spatial and temporal incoherence. (a) Spatially incoherent sources are comprised of many spatial modes, each of which generates a separate uncorrelated speckle realization. As the modes have no phase correlation at long detection times, their intensities are superimposed, and the speckle contrast is efficiently suppressed. (b) Temporally incoherent sources are comprised of many temporal modes (different wavelengths), whose intensities are superimposed incoherently on the measuring device, at long detection times. For temporal modes with the same transverse structure (same spatial mode), the difference in wavelength between modes results in radial scaling of the generated speckle pattern, generally leading to limited reduction of speckle contrast.}
	\label{fig:speckle1}
\end{figure*}

In order to reduce the spatial coherence, and thereby speckle noise, several methods have been developed. These commonly involve  mechanical or electro-optical methods~\cite{Bromberg2014a, Wang2013a, Turunen1991, Lohmann1999, MCKECHNIE197535, KAWAGOE1982197, Akram:10, ambar1986fringe, saloma1990speckle, dingel1992laser,dingel1993speckle, imai1980optical, Waller2012} for generating many uncorrelated speckle realizations from the same object and then summing their intensities incoherently. One of the simplest and most common methods for speckle suppression is achieved by a moving diffuser that is placed between the laser source and the object~\cite{Goodman2005a}. As these methods are based on accumulation of a large series of independent speckle realizations, they are time consuming and require relatively long integration time (i.e. exposure time, the fraction of time over which the detector is exposed), and are therefore inadequate for high speed imaging applications. An alternative method for speckle suppression is based on broad bandwidth low temporal coherence sources, such as supercontinuum sources and superluminescent diodes. However these sources are spatially coherent and suffer from non-negligible speckle noise in many imaging systems~\cite{Redding2012}, as illustrated in Fig.~\ref{fig:speckle1}. In fact, except for special non-generic cases that combine very broad bandwidth, thick scattering media and large scattering angles, spatial incoherence (rather than temporal incoherence) is essential for significant speckle contrast suppression~\cite{Goodman2005a, Redding2012}.

Recently, several novel multimode lasers with inherently low spatial coherence have been developed~\cite{Redding2012, Nixon2013g, Knitter2016, Redding2015b, Redding2014, Hokr2016a}. With these, each spatial mode generates an independent uncorrelated speckle realization, and consequently reduces the speckle contrast. Specifically, for a laser of $N$ spatial modes, the total intensity is the incoherent sum of the intensities of all spatial modes, and results in the well known relation of the speckle contrast, $C=N^{-1/2}$~\cite{Nixon2013g}. 

Notice however that this relation is true only in the regime of long time scales. In general, the speckle contrast is expected to vary with detection time of the measuring device, and to depend on the temporal dynamics of the laser source. To see this intuitively, consider for example the light of $N$ spatial modes, which are separated by some frequency spacing $\Delta\nu$, illuminating a diffuser. Each mode will generate an independent uncorrelated speckle realization. For long exposure time $\Delta\tau\gg\Delta\nu^{-1}$, the total intensity is an incoherent sum of the intensities of all modes and the speckle contrast is $C=N^{-1/2}$, as noted above. For very short exposure time $\Delta\tau\ll\Delta\nu^{-1}$, the light of spatial modes interfere with one another and generate a new speckle realization with speckle contrast of $C=1$. This example illustrates that speckle contrast depends on exposure time. While the reduction of spatial coherence (and accordingly of speckle contrast) in highly multimode lasers is well established in the regime of long time scales, the theoretical framework of spatial coherence reduction in short time scales is limited. 

In this paper we concentrate on short time scales, typically in the nanosecond regime, and present a new spatio-temporal mechanism of \textit{spatial} coherence reduction, in which the \textit{temporal} (longitudinal) modes play a dominant role. In this regime of exposure times and for lasers with a large number of spatial modes ($N\gg M$, $N$ and $M$ being the number of spatial and temporal modes, respectively), we show experimentally and theoretically that speckle contrast depends only on the number of longitudinal modes, and not on the number of spatial modes. This dependence is true even though the bandwidth of the laser is narrow compared to its central frequency, indicating that the reduction of speckle contrast if caused by spatial rather than temporal incoherence. We show that spatial modes that are close in frequency behave as supermodes and contribute to speckle reduction in short time scales, on the order of the inverse of the laser's free spectral range (FSR). The nature of the supermodes is studied numerically, and their non-localized effect on the spatial coherence is described. Explicitly, for the short time scale regime, the speckle contrast of a laser with $M$ longitudinal modes is $C=M^{-1/2}$. In this regime, the functional form of the spatial coherence has a bimodal distribution, which consists of a narrow peak and uniform non-zero background. The width of the spatial coherence is determined by the number of spatial modes, and its overall background decays as $M^{-1/2}$. In long time scales, the speckle contrast has the traditional relation of $C=N^{-1/2}$ and the background of the spatial coherence functional form reduces to zero.

Speckles are involved in fundamental research~\cite{Bromberg2014a, Wang2011a, Schwartz2007a, Billy2008a, Bromberg2010a} and in many applications~\cite{Fercher1985a, Ferri2005a, Boas2010a, Redding2013a}, so a better understanding of their temporal dynamics and suppression are of great interest. For example, speckle suppression for short exposure times can be advantageous for material processing~\cite{dickey2017laser} and ultra-fast full-field imaging applications, such as flow dynamic measurements in aerodynamics\cite{Dolan2004}, and study of shock-waves and ablation~\cite{Tsuboi1994}. In addition, there is a growing need for rapid speckle suppression in trapping of ultra-cold atoms in large optical lattices, which are often used as quantum simulators~\cite{Barredo2016, endres2016atom, gross2017quantum}. Such large optical lattices are exceptionally sensitive, and are currently limited by speckle noise that arises from random scattering events. Trapping of ultra-cold atoms requires extremely effective speckle suppression down to a level of $\sim 100$dB, at time scales that are dictated by the trapping frequency of the optical lattice, typically sub microsecond. Such strenuous requirements are difficult to achieve with mechanical and electro-optical methods. \newpage
\section{II. Experimental arrangement\newline
and results}
As noted above, speckle contrast in the regime of long exposure times depends on the number of spatial modes, as $N^{-1/2}$. So, if significant speckle suppression is required (i.e. speckle contrast of few percent), a highly incoherent laser source with hundreds or even thousands of spatial modes is needed. For such a source we resorted to a degenerate cavity laser (DCL)~\cite{Arnaud1969,Siegman1986} and investigated its temporal dynamics of speckle suppression. 

The DCLs have the general property that any diffraction limited spot is accurately imaged onto itself after a single round-trip. Consequently, these cavities can support any arbitrary transverse field distribution, a property which has been used for different applications, such as enhanced imaging~\cite{hardy1965active, Nixon2013c, nimmrichter2017full}, laser scanning~\cite{myers1967electron}, phase locking of laser arrays~\cite{he2006phase,ronen2011phase, Nixon2013b, pal2017observing} and simulating physical systems~\cite{Nixon2013b, zhou2017dynamically}. Since all modes are degenerate in loss, they can all lase simultaneously, despite mode competition. This property has been recently exploited for spatial coherence control and efficient speckle suppression~\cite{Nixon2013g, Chriki2015, Knitter2016}. 
 
\begin{figure}
    \includegraphics{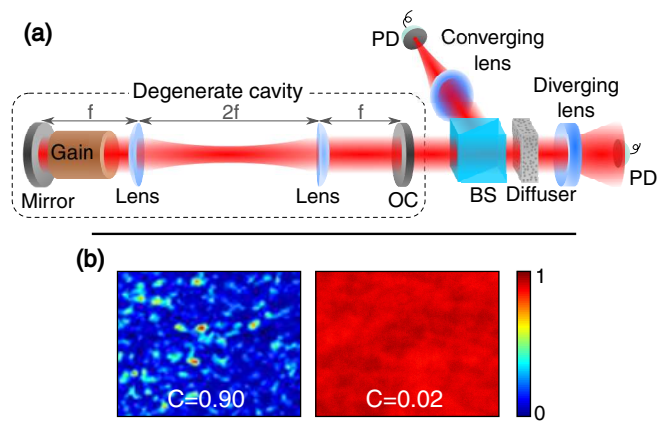}
	\caption{Laser and detection arrangements, and representative speckle images. (a) Experimental arrangements of the DCL and of the speckle contrast detection. OC – output coupler; BS – beam splitter; PD – photodetector. (b) Representative speckle images with a spatially coherent source (left) and with the DCL (right) for long exposure times.}
	\label{fig:experiment}
\end{figure}

In our experiment, the DCL was comprised of a high reflective mirror, Nd:YAG gain medium, two lenses with $f=25\cm$ in a $4f$ telescope configuration and an output coupler, as shown Fig.~\ref{fig:experiment}(a). The gain medium was pumped by a double Xenon flash lamp with quasi-CW $100\mu\sec$ long pulses, and the reflectivity of the output coupler was $R=80\%$. The $4f$ telescope inside the cavity assured that any field distribution is accurately imaged onto itself after every round-trip, so any field distribution is a degenerate eigenmode of the cavity. The diameter of the gain medium was $0.95\cm$, much larger than the diffraction spot size of the telescope, so the cavity supported a huge number of spatial modes ($N=320,000$~\cite{Nixon2013g}), i.e. it was a highly spatially incoherent source. Figure~\ref{fig:experiment}(b) shows two representative speckle images for long exposure times, one obtained with a spatially coherent Nd:YAG laser and the other with the DCL. As evident, the speckle contrast is greatly reduced with the DCL, indicating that the DCL indeed supports a very large number of spatial modes.

The number of lasing longitudinal modes in the DCL was determined by the ratio between the total bandwidth of the laser (measured by means of a Michelson-interferometer as $32\GHz$) and the free spectral range of the cavity (measured from the beating frequencies in the laser output as $123\MHz$), to yield $M=260$~\cite{Supplementary}. Our laser thus fulfills the condition $N\gg M\gg 1$.

To measure temporal dynamics of the speckle contrast from the DCL, a thin optical diffuser was placed outside the cavity, and a fast photodetector, much smaller than the typical speckle size, measured the intensity time dynamics of a single point in the speckle field, as shown in the right part of Fig.~\ref{fig:experiment}(a). The speckle contrast was measured by performing 1000 uncorrelated time series measurements with a fixed rotation of the diffuser (and hence rotation of the speckle field) before each measurement. The speckle field measurements were normalized by the total output intensity, measured by a second fast photodetector as shown in the right part of Fig.~\ref{fig:experiment}(a).

\begin{figure}
    \includegraphics{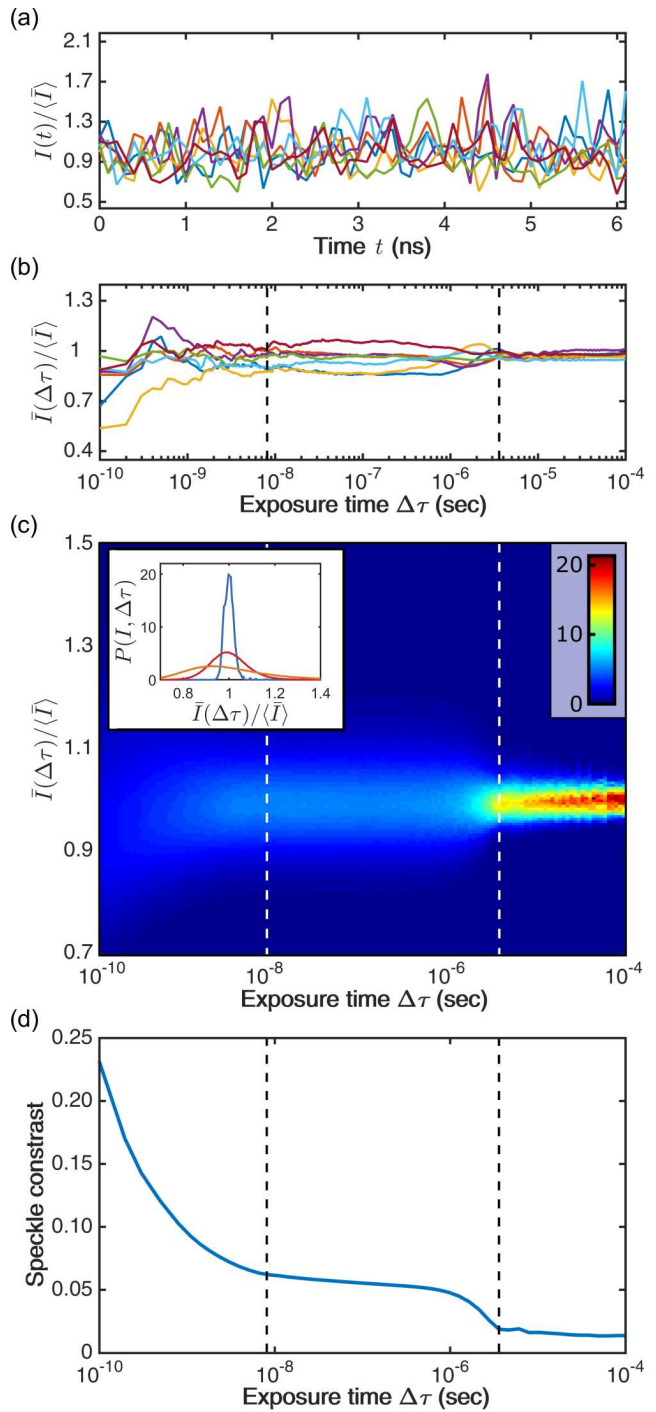}
	\caption{Experimental temporal dynamics of speckles. (a) Intensity time series of seven randomly selected points in the speckles field. (b) Corresponding average intensity as a function of exposure time for the selected seven points. (c) Intensity probability distribution as a function of exposure time, calculated from 1000 uncorrelated time series measurements. Inset shows crossections of this image for $\Delta\tau=10^{-3} \sec$ (blue), $\Delta\tau=10^{-7} \sec$ (red), $\Delta\tau=10^{-10} \sec$ (orange). (d) Speckle contrast as a function of exposure time, calculated from 1000 uncorrelated time series measurements. Dashed lines in (b-d) denote transitions between exposure time regimes.}
	\label{fig:ExpData}
\end{figure}

The experimental results of speckle time dynamics are presented in Fig.~\ref{fig:ExpData}. Figure~\ref{fig:ExpData}(a) shows representative time series measurements for seven randomly selected points in the speckle field. Figure~\ref{fig:ExpData}(b) shows the corresponding seven time average intensities $\bar{I}(\Delta\tau)=\frac{1}{\Delta\tau}\int_0^{\Delta\tau} I(t)dt$ simulating the effect of finite exposure time $\Delta\tau$ . Based on histograms of many such intensity measurements, we determined the intensity probability distribution $P(\bar{I}, \Delta\tau)$ as a function of exposure time, shown in Fig.~\ref{fig:ExpData}(c). As evident, the width of the probability distribution decreases at long exposure time. To obtain more insight, we calculated the dependence of speckle contrast $C(\Delta\tau)$ on exposure times, using $C(\Delta\tau)=\sqrt{\langle\bar{I}(\Delta\tau)^2\rangle)/\langle\bar{I}(\Delta\tau)\rangle^{2}-1}$, where $\langle\cdot\rangle$ denotes an ensemble average over 1000 different points in the speckle field. For long exposure times, the speckle contrast measurements were also performed conventionally by full-field imaging of the speckle field with a CMOS camera, confirming the validity of our method for measuring speckle contrast (see Supplemental Material~\cite{Supplementary}).

The resulting speckle contrasts as a function of exposure times are presented in Fig.~\ref{fig:ExpData}(d). Three temporal regimes are clearly identified: (a) long exposure time regime ($\Delta\tau>\SI{3}{\mu\sec}$) where the speckle contrast is reduced to $C=0.014$, slightly higher than that expected from measurements of the beam quality factor (not shown)~\cite{Nixon2013g},  i.e. $N^{-1/2}=0.002$, due to non-uniformity of the illuminating beam; (b) short exposure time regime ($\SI{10}{\nsec}<\Delta\tau<\SI{1}{\mu\sec}$), where speckle contrast is $C\sim 0.056$, in good agreement with the expected value $M^{-1/2}=0.062$ (see next section); and (c) instantaneous exposure time regime ($\Delta\tau<\SI{1}{\nsec}$), where the speckle contrast increases inversely with exposure time, and is expected to reach its maximal value of $C=1$ for exposure time shorter than 31ps (the inverse of the lasing spectral bandwidth of $32\GHz$). Such instantaneous time scale was beyond the resolution of our photodetector. These three temporal regimes are also evident from the intensity probability distribution $P(I, \Delta\tau)$ of Fig.~\ref{fig:ExpData}(c). Transitions between the different exposure time regimes are denoted by dashed lines, and they correspond to the calculated and measured free spectral range ($123\MHz$), and to the expected minimal broadening of a single mode due to mechanical instabilities ($\sim 300\kHz$)~\cite{Koechner2003}. 

We repeated the measurements of speckle contrast as a function of exposure time for lasing with other numbers of spatial modes, and found similar results with the same three temporal regimes for all measurements (see Supplemental Material~\cite{Supplementary}).


\section{III. Modal frequency spectrum analysis}

The temporal dynamics of the speckle contrast $C$ are strongly related to the spectral frequencies of the laser. To calculate these frequencies we resort to Gaussian mode theory, and describe the propagation of the field in each round-trip using a ray transfer matrix representation. A ray transfer matrix is defined such that any point $x_0$ near the output with ray slop $\dot x_0$, will propagate after one round-trip to a point $x_1$ with ray slope $\dot x_1$ according to
\begin{equation}
        \begin{pmatrix}
        x_1 \\ \dot x_1
        \end{pmatrix}
        =
        \begin{pmatrix}
            A & B \\ C & D
        \end{pmatrix}
        \begin{pmatrix}
            x_0 \\ \dot x_0
        \end{pmatrix},
\end{equation}
where $A$, $B$, $C$, $D$ are the ray transfer matrix elements. 

\begin{figure}
    \includegraphics{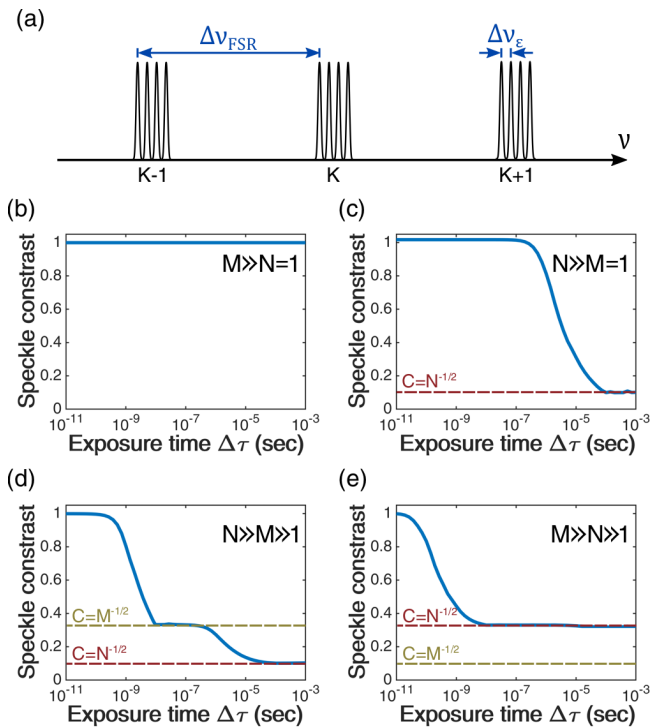}
	\caption{Numerical calculation results of speckle contrast. (a) Part of modal frequency spectrum assumed in the numerical calculations. (b-e) Calculated speckle contrast as a function of exposure time for laser cavity with (b) $M=100$ longitudinal modes and $N=1$ spatial modes, (c) $M=1$ and $N=100$, (d) $M=10$ and $N=100$, and (e) $M=100$ and $N=10$.
}
	\label{fig:SimSpeckle}
\end{figure}

Based on this formalism, it can be shown that the frequency $\nu_{q,m,n}$ for the $q^{\text{th}}$ longitudinal mode and spatial mode of order $(m, n)$ is~\cite{Arnaud1969, arai2013accumulated}
\begin{equation}
\label{eq:nu_qmn1}
		\nu_{q,m,n}=\frac{c}{2L} \left[ q+(m+n+1)cos^{-1} \left( \sqrt{\frac{A+D}{4}+\frac{1}{2}} \right) \right]
\end{equation}
where $c$ is the speed of light and $L$ is the total length of the cavity, and therefore $c/2L$ is the FSR of the cavity. For an ideal degenerate cavity $A=D=1$ and $B=C=0$, resulting in prefect degeneracy of all modes: the allowed frequencies  are independent of $m$ and $n$. Aberrations in the cavity generally break this degeneracy. For example, isotropic aberrations would result in~\cite{Supplementary}
\begin{equation}
\label{eq:nu_qmn2}
		\nu_{q,m,n}=\frac{c}{8f}[q+\epsilon(m+n+1)]
\end{equation}
where $f$ is the focal distance of the lenses in the cavity ($L=4f$), as mentioned above, and $\epsilon$ depends on the type and degree of the aberrations. For the realistic case of low aberrations $\epsilon\ll 1$, the frequency spacing $\Delta\nu_\epsilon$ between modes with the same value of $K$ is much smaller than the FSR $\Delta\nu_{FSR}$, as illustrated in Fig.~\ref{fig:SimSpeckle}(a). More accurately, it is defined as the spectral bandwidth of a single mode or the frequency spacing between adjacent modes,  whichever is larger.  

The modal frequency spectrum in Fig.~\ref{fig:SimSpeckle}(a) is applicable to a broad family of cavities that are mode degenerate. These include several important laser cavities, such as the planar mirror (Fabry-Perot) laser cavity and the near concentric laser cavity\cite{Siegman1986}. The modal frequency spectrum with separation of time scales is also similar to that of multimode fiber lasers that are not extremely long (typical length$\lesssim100\m$). We therefore see a general link between the near degeneracy of the spatial modes in gain and loss (enabling support of many spatial modes even in the presence of mode competition~\cite{Nixon2013g, Chriki2015}) and their near degeneracy in frequency. 

Based on this modal frequency spectrum we performed numerical calculations to determine the speckle contrast as a function of exposure time and number of spatial and longitudinal modes. The frequency spacing between modes were taken to be $\Delta\nu_\epsilon=10\kHz$ and $\Delta\nu_{FSR}=100\MHz$, i.e. $\Delta\nu_\epsilon\ll\Delta\nu_{FSR}$ as in our experiments. Figures~\ref{fig:SimSpeckle}(b) and~\ref{fig:SimSpeckle}(c) present results for the simple cases of $M\gg N=1$ and $N\gg M=1$, respectively, while Fig.~\ref{fig:SimSpeckle}(d) presents the results for the case of $N\gg M\gg 1$ that corresponds to our experiment. Figure~\ref{fig:SimSpeckle}(b) shows the calculated speckle contrast as a function of exposure time for a laser with $M=100$ longitudinal modes and a single spatial mode $N=1$. As evident, the speckle contrast is equal to unity for all exposure times of the detecting device. This demonstrates that spatial diversity is a dominant component in speckle suppression with typical laser sources and thin scattering media. Figure~\ref{fig:SimSpeckle}(c) shows the calculated speckle contrast as a function of exposure time for a laser with a single longitudinal mode, $M=1$, and $N=100$ closely spaced spatial modes. As evident, the speckle contrast is suppressed to $C=N^{-1/2}=0.1$, but only for $\Delta\tau\gg\Delta\nu_\epsilon^{-1}$, when all the spatial modes are mutually incoherent. For $\Delta\tau\ll\Delta\nu_\epsilon^{-1}$, the spatial modes interfere with one another, and generate a new speckle pattern with speckle contrast of unity.

Figure~\ref{fig:SimSpeckle}(d) shows the calculated speckle contrast as a function of exposure time for a laser with $N=100$ spatial modes and $M=10$ longitudinal modes. Three distinct temporal regimes are clearly seen, in agreement with the experimental data in Fig.~\ref{fig:ExpData}(c). For long exposure times $\Delta\tau\gg\Delta\nu_\epsilon^{-1}$, the total intensity distribution is an incoherent sum of all spatial modes, whereby the speckle contrast is 
$C=N^{-1/2}=0.1$. For instantaneous exposure times $\Delta\tau\ll\Delta\nu_{FSR}^{-1}$, all spatial modes interfere with one another and the speckle contrast reaches unity. In the regime of short exposure times, $\Delta\nu_{FSR}^{-1}\ll\Delta\tau\ll\Delta_\epsilon^{-1}$, the speckle contrast is $C=M^{-1/2}=0.32$. 

\begin{figure*}
    \includegraphics{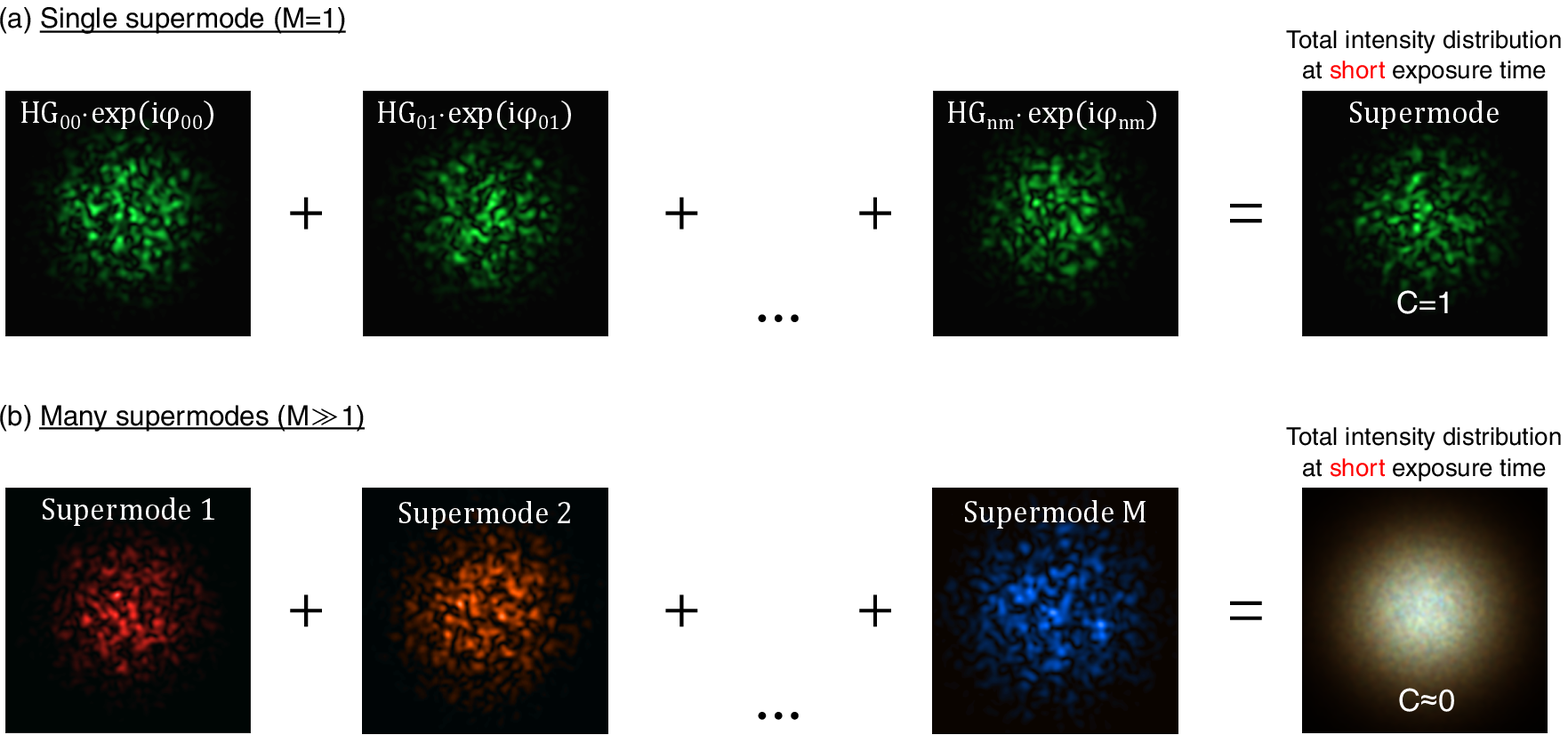}
	\caption{Illustrations diagram of the spatio-temporal mechanism that leads to speckle suppression at short time scales. (a) All spatial modes in the vicinity of a given frequency $qc/8f$ interfere coherently to generate a single speckle pattern with speckle contrast of unity. Note each spatial mode has an independent random phase, which is determined by the random initial lasing process, and is constant in the short time scale $\Delta\tau\ll\Delta\nu_\epsilon^{-1}$. (b) The supermodes are mostly uncorrelated with one another, and interfere incoherently in the time scale                                                                                    $\Delta\tau\gg\Delta_{FSR}$, thereby suppressing speckle contrast.
}
	\label{fig:Speckle2}
\end{figure*}

We propose the following spatio-temporal mechanism to explain the surprising dependence of $C$ on $M$ at the intermediate regime of short time scales. For  $\Delta\tau\ll\Delta\nu_\epsilon^{-1}$, all spatial modes that lie in the vicinity of a given frequency $qc/8f$ interfere coherently to generate a speckle pattern with speckle contrast of unity, and can therefore be considered as a single supermode, as illustrated in Fig.~\ref{fig:Speckle2}(a). However, due to the random process of lasing, a random phase is assigned to each spatial mode, and consequently each supermode generates a different speckle realization. The degree of orthogonality between supermodes is determined by the values of these random phases, and can be considered as a random walk in the complex plane. Consequently, the supermodes are in fact only quasi-modes, and the speckle realizations of different supermodes are partially correlated. In the short time regime we are considering, $\Delta\tau$ is large compared to $\Delta\nu_{FSR}^{-1}$, and therefore the supermodes do not interfere and their intensities are accumulated incoherently, as illustrated in Fig.~\ref{fig:Speckle2}(b). Considering the partial correlations between supermodes, it can be shown that the speckle contrast is (see Supplemental Material~\cite{Supplementary})
\begin{equation}
\label{eq:CvsMN}
    C=\frac{1}{\sqrt{N[1-(1-1/N)^M ]}}.
\end{equation}
In the limit $N\gg M$, Eq.~\ref{eq:CvsMN} reduces to $C=M^{-1/2}$, namely, the speckle contrast is suppressed as the number of longitudinal modes, in agreement with the experimental and numerical data of Figs.~\ref{fig:ExpData}(d) and~\ref{fig:SimSpeckle}(d).

Figure.~\ref{fig:SimSpeckle}(e) shows the calculated speckle contrast as a function of exposure time for a laser with $N=10$ spatial modes and $M=100$ longitudinal modes. As evident, the speckle contrast is only suppressed to a level of $C=0.32=N^{-1/2}$. This corresponds to $N\ll M$, where Eq.~\ref{eq:CvsMN} reduces to $C=N^{-1/2}$, indicating that the speckle contrast cannot be reduced by more than the number of spatial modes, and reflecting the essential role of spatial diversity for effective speckle suppression. 

\section{IV. Dependence of spatial coherence on exposure time}

We also investigated analytically and numerically the effect of exposure time on the spatial coherence. We started with the spatial field correlations (mutual intensity), 
\begin{equation}
\label{eq:mu}
    J(\mathbf{x}_1, \mathbf{x}_2; \Delta\tau)=\langle E(\mathbf{x}_1;t)E^*(\mathbf{x}_2;t)\rangle_{\Delta\tau}, 
\end{equation}
where $\mathbf{x}_1$ and $\mathbf{x}_2$ are 2D transverse coordinates, $E(\mathbf{x};t)$ is the electric field at point $\mathbf{x}$ and time $t$, and $\langle\cdot\rangle_{\Delta\tau}$ denotes temporal averaging, 
\begin{equation}
\label{eq:mu2}
    \begin{split}
        \langle f(\mathbf{x}_1,\mathbf{x}_2&;t)\rangle_{\Delta\tau}\equiv\\
        & \frac{1}{T\Delta\tau} \int_{t=0}^T\left|\int_{t'=t}^{t+\Delta\tau}f(\mathbf{x}_1,\mathbf{x}_2; t')dt'\right|dt.
    \end{split}
\end{equation}
According to the Van Cittert Zernike theorem, the modulus of the mutual intensity $|J|$ for a spatially incoherent source depends only on the the difference of coordinates, so $|J(\mathbf{x}_1,\mathbf{x}_2;\Delta\tau)|=|J(\Delta\mathbf{x};\Delta\tau)|$, where $\Delta\mathbf{x}\equiv\mathbf{x}_2-\mathbf{x}_1$~\cite{Goodman2015}. It is often convenient to deal with the normalized version of $J(\mathbf{x}_1, \mathbf{x}_2; \Delta\tau)$, 
\begin{equation}
\label{eq:define_mu}
    \mu(\mathbf{x}_1,\mathbf{x}_2;\Delta\tau)=\frac{J(\mathbf{x}_1,\mathbf{x}_2;\Delta\tau)}{\sqrt{I(\mathbf{x_1}; \Delta\tau)I(\mathbf{x}_2; \Delta\tau)}},    
\end{equation}
where $\mu(\mathbf{x}_1,\mathbf{x}_2;\Delta\tau)$ is often referred to as the complex coherence factor.

The field at the output of the DCL is a superposition of all spatial and longitudinal modes, and can therefore be expressed as 
\begin{equation}
\label{eq:Eout}
    E_{out}(\mathbf{x})=\sum_m\sum_n E_{mn}(\mathbf{x})e^{i(\omega_{mn}t+\phi_{mn})},
\end{equation}
where $E_{mn}$ is a mode with longitudinal mode index $m$ and spatial mode index $n$, $\omega_{mn}$ is the frequency of that mode, and $\phi_{mn}$ is its random phase. In the long exposure time regime, the mutual intensity $J(\Delta\mathbf{x}; \Delta\tau\gg\Delta\nu_\epsilon^{-1})$ can be readily obtained, as 
\begin{equation}
\label{eq:mu_long}
        J(\Delta\mathbf{x}; \Delta\tau\gg\Delta\nu_\epsilon^{-1}) = \sum_m \sum_n E_{mn}^{}(\mathbf{x}_0)E_{mn}^*(\mathbf{x}_0+\Delta\mathbf{x}),
\end{equation}
where $\mathbf{x}_0$ is any arbitrarily selected point. Due to the summation of orthogonal modes, the resulting $|J|$ has narrow peaked distribution, with width $d_c$ that scales as $d_c\sim D/N$, where $D$ is the transverse size of the source. 

Similarly, in the short exposure time regime, the mutual intensity $J(\Delta\mathbf{x}; \Delta\nu_{FSR}^{-1}\ll\Delta\tau\ll\Delta\nu_\epsilon^{-1})$ is
\begin{equation}
\label{eq:mu_short}
    \begin{split}
        J&(\Delta\mathbf{x}; 
        \Delta\nu_{FSR}^{-1}\ll\Delta\tau\ll\Delta\nu_\epsilon^{-1}) = \\ 
        &\sum_m \sum_{n=n'}E_{mn}^{}(\mathbf{x}_0)E_{mn'}^*(\mathbf{x}_0+\Delta\mathbf{x}) + \\
        &\sum_m \sum_{n\ne n'}E_{mn}^{} (\mathbf{x}_0)E_{mn'}^*(\mathbf{x}_0+\Delta\mathbf{x})e^{i(\phi_{mn}-\phi_{mn'})}. 
    \end{split}
\end{equation}
Notice the first term is equal to the mutual coherence at long exposure time $J(\Delta\mathbf{x}_0, \Delta\tau\gg\Delta\nu_\epsilon^{-1})$, and accordingly it has a narrow peaked distribution, where the area at full width half maximum (FWHM) is inversely proportional to the number of spatial modes $N$. The second term adds random fluctuations with average value that decays as $M^{-1/2}$, due to the random phases that enter the sum. Therefore we conclude it is the background of $|J|$ which invokes the increase in speckle contrast in the short exposure time regime, as compared to that in the long exposure time regime. 

\begin{figure}
    \includegraphics{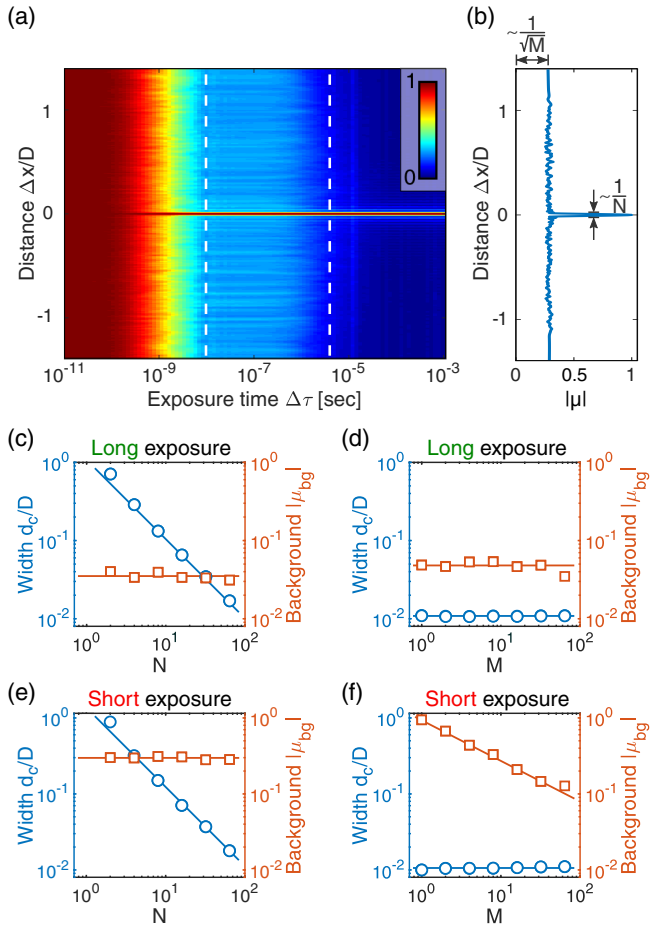}
	\caption{Calculated modulus of the complex (spatial) coherence factor $|\mu|$. (a) Modulus $|\mu|$ as a function of exposure time $\Delta\tau$ and normalized difference of coordinates $\Delta\mathbf{x}/D$. The calculations were performed for $N=100$ spatial modes and $M=10$ longitudinal modes. (b) Corresponding crossection, showing the modulus $|\mu|$ as a function of the normalized difference of coordinates, for short exposure time of $\Delta\tau=10^{-7} \sec$. (c) Normalized width $d_c/D$ and background $|\mu_{bg}|$ of the spatial coherence as a function of the number of spatial modes $N$ in long exposure time of $\Delta\tau=10^{-3}\sec$, for $M=10$ longitudinal modes. (d) Normalized width $d_c/D$ and background $|\mu_{bg}|$ as a function of the number of longitudinal modes $M$ in long exposure time of $\Delta\tau=10^{-3}\sec$, for $N=100$ spatial modes. (e) Normalized width $d_c/D$ and background $|\mu_{bg}|$ as a function of the number of spatial modes $N$ in short exposure time of $\Delta\tau=10^{-7}\sec$, for $M=10$ longitudinal modes. (f) Normalized width $d_c/D$ and background $|\mu_{bg}|$ as a function of the number of longitudinal modes $M$ in short exposure time of $\Delta\tau=10^{-7}\sec$, for $N=100$ longitudinal modes. (a-f) were obtained by averaging 100 realizations.
}
	\label{fig:SimSpatial}
\end{figure}

To verify our analysis of the spatial coherence in the long and short exposure time regimes, we carried out detailed numerical simulations, based on the model presented above. The results are presented in Fig.~\ref{fig:SimSpatial}. Figure \ref{fig:SimSpatial}(a) presents the calculated modulus of the complex coherence factor $|\mu(\Delta\mathbf{x};\Delta\tau)|$ as a function of transverse distance $\Delta\mathbf{x}$ and exposure time $\Delta\tau$, for $N=100$ and $M=10$ in 1D geometry, and with $\Delta\nu_\epsilon=10\kHz$ and $\Delta\nu_{FSR}=100\MHz$. Again, as for the intensity probability distribution and speckle contrast, three temporal regimes are observed: in the long exposure time regime, $\Delta\tau\gg\Delta\nu_\epsilon^{-1}$, $|\mu|$ is narrow peaked with low background; in the short time regime, $\Delta\nu_{FSR}^{-1}\ll\Delta\tau\ll\Delta\nu_\epsilon^{-1}$, $|\mu|$ is narrow peaked, but with significant background, as clearly evident in Fig.~\ref{fig:SimSpatial}(b); and in the instantaneous time regime, $\Delta\tau\ll\Delta\nu_{FSR}^{-1}$, $|\mu|$ is uniformly equal to unity everywhere in space. As predicted in our analysis, the width of the central peak of $|\mu|$ is the same in the short exposure time regime as that in the long exposure time regime, but the level of background is different. 

We also numerically investigated the dependence of the width and background of $|\mu|$ on the number of spatial and temporal modes, $N$ and $M$, in both the short and long exposure time regimes. The results are presented in Figs.~\ref{fig:SimSpatial}(c-f). Figures~\ref{fig:SimSpatial}(c-d) show the results in the long exposure time regime. Here the width of spatial coherence $d_c$ decays with the number of spatial modes, and is independent of the number of longitudinal modes. The level of background of $|\mu|$ is fixed at a negligible level (limited in our simulations to numerical artifacts), and does not vary with the number of spatial or longitudinal modes (except for the special case of $N=1$ where the spatial coherence is trivially equal to one). These properties of $|\mu|$ can be readily understood in the long exposure time regime. In this regime, the number of longitudinal modes is not expected to affect the spatial coherence properties of the laser; and more lasing spatial modes (larger $N$) indicates smaller transverse regions of spatial coherence (smaller $d_c$). Figures~\ref{fig:SimSpatial}(e-f) show the results in the short exposure time regime. Here the width $d_c$ depends on the number of spatial modes $N$, as in the regime of long the exposure time. However, the level of background $|\mu_{bg}|$ in this temporal regime depends only on the number of longitudinal modes $M$. Specifically, the level of background decays as $|\mu_{bg}|\sim M^{-1/2}$, in full agreement with our analysis above.

In previous sections, we found that the suppression of speckles in the short and long time regimes has similar mathematical form, namely that the speckle contrast decays as the square root of the number of degrees of freedom, $N$ or $M$. Here we see that the physical mechanism responsible for speckle suppression in the two regimes is quite different. While the reduction of speckle contrast in long exposure times is caused by narrow local regions of coherence, the reduction of speckle contrast in short exposure times is dominated by global reduction of the spatial coherence. This, of course, is related to the non-localized nature of the  spatio-temporal supermodes, which are responsible for speckle suppression in short time scales.

\section{V. Imaging of a fast moving object\newline in the short exposure time regime}

\begin{figure}
    \includegraphics{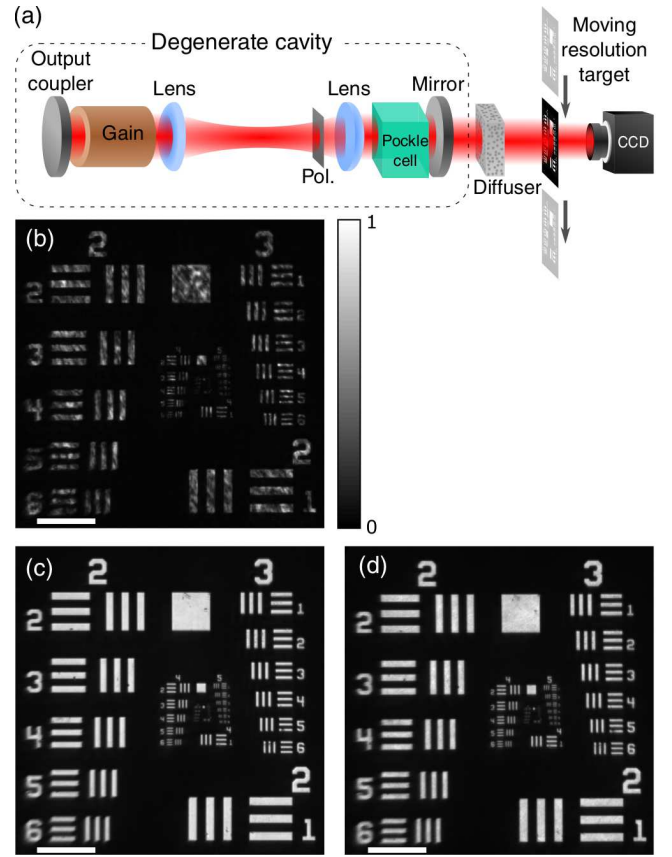}
	\caption{Experimental full-field imaging results of an Air Force resolution target. (a) Experimental arrangement. (b) Detected image of a static object using a spatially coherent source and long exposure time. (c) Detected image of a static object using a spatially incoherent source and long exposure time. (d) Detected image of a fast moving object and short exposure time. Scale bars, $500 \mu m$.}
	\label{fig:MovingDiff}
\end{figure}

Rapid reduction of the speckle contrast is extremely relevant and desired to many ultra-fast full-field imaging applications. To demonstrate this, we performed short exposure time full-field imaging of a fast moving object, and compared our results to those obtained with a stationary object and long exposure time. In the experiments, we used the arrangement shown schematically in Fig.~\ref{fig:MovingDiff}(a). As shown, the output of the laser illuminated a fast moving U.S. Air Force resolution target, after propagating through a transparent optical diffuser (Newport $10^o$ light shaping diffuser). The target was rotated rapidly such that it had a linear velocity of $15\m/\sec$. A Pockles cell was placed inside the cavity, and the DCL was actively Q-switched. The resulted   short $100\nsec$ output pulses of the DCL laser limited the effective exposure time of the camera, and enabled high spatial resolution of the fast moving Air Force target, since the  target could only move $1.5\mu m$ in each pulse.

Figures~\ref{fig:MovingDiff}(b-d) present the experimentally detected images. Figure~\ref{fig:MovingDiff}(b) shows the detected image of a static object using a coherent source. As evident, the image suffers from high speckle contrast. Figure~\ref{fig:MovingDiff}(c) shows the detected image of a static object that is illuminated with spatially incoherent light from the DCL and long exposure time. As evident, the speckle contrast is suppressed, and was measured to be $C=0.015$. Figure~\ref{fig:MovingDiff}(d) shows the detected image of the fast moving object with short exposure time. Here again the speckle contrast is suppressed, and was measured to be $C=0.045$, while the spatial resolution remains high due to the short exposure time. These results demonstrate that speckle suppression can indeed be obtained at short nanosecond time scales.

For direct comparison, consider speckle suppression by means of a rotating ground glass diffuser and a Q-switched single mode laser. The speed of commercially available rotating stages are typically offer $0.1-3\Hz$, and a typical grain size of a ground glass diffusers can be as small as $\sim 2\mu m$. Therefore, new speckle realizations can occur every $1-100\mu sec$ - considerably longer than the Q-switched pulse duration. Even if an extremely fast rotating diffuser is used, it wouldn't be possible to accumulate more than just few speckle realizations during the $100\nsec$ pulse duration, and only slight speckle suppression would be achieved, if at all.


\section{VI. Concluding remarks}
We presented and characterized a new spatio-temporal mechanism of laser sources that influences the spatial coherence at short time scales. Counter intuitively, we showed that in the regime of short time scales, spatial coherence, and consequently speckle noise, are influenced by the number of longitudinal modes, and not only by spatial modes. Specifically, for a degenerate cavity with $N\gg M\gg1$, we identified, both in theory and experiment, three distinct temporal regimes. In the regime of long time scales of $\Delta\tau\gg\Delta\nu_\epsilon^{-1}$, the speckle contrast is $C=N^{-1/2}$. In the regime of short time scales of $\Delta\nu_{FSR}^{-1}\ll\Delta\tau\ll\Delta\nu_\epsilon^{-1}$, longitudinal modes in the vicinity of one common frequency form  independent quasi-supermodes, and the speckle contrast is governed by the number of longitudinal modes supported by the laser, as $C=M^{-1/2}$. In the instantaneous very short time scale regime of $\Delta\tau\ll\Delta\nu_{FSR}^{-1}$, the speckle contrast is trivially equal to unity, $C=1$. 

The new spatio-temporal mechanism was attributed to the presence of non-localized spatio-temporal supermodes, which suppress the overall background of the spatial coherence functional form. It can be applied to various different types of lasers, such as to highly multimode fiber lasers in order to suppress speckles even in the sub-picosecond time scale, and it can enable control of the speckle contrast of short time scales by changing the number of longitudinal modes in the cavity. Such speckle suppression and control in short time scales can be advantageously exploited in material processing, optical trapping of ultra-cold atoms and ultra-fast imaging, as was exemplified in a speckle-free full-field imaging of a fast moving object.

\begin{acknowledgments}
We thank Prof. Hui Cao (Yale University) for many fruitful discussions, and Ophir Turetz and Alexander Cheplev for their help in spectral measurements. This work was supported in part by the Israel Science Foundation and the Israel-US Binational Science foundation.
\end{acknowledgments}

\bibliographystyle{apsrev4-1}
\bibliography{library}


\pagebreak
\clearpage
\widetext
\begin{center}
\textbf{\large Supplemental Material for\\ 
Spatio-Temporal Supermodes: Rapid Reduction of Spatial Coherence\\ in Highly Multimode Lasers}

\bigskip
Ronen Chriki,$^\text{1}$ Simon Mahler,$^\text{1,2}$ Chene Tradonsky,$^\text{1}$ Vishwa Pal,$^\text{1}$ Asher A. Friesem,$^\text{1}$ and Nir Davidson$^\text{1}$

\smallskip
$^\text{1}$ \textit{Department of Physics of Complex Systems, Weizmann Institute of Science, Rehovot 7610001, Israel}

$^\text{2}$ \textit{Univ. Paris Sud, Universit\'e Paris Saclay, 91405 Orsay, France}
\end{center}

\setcounter{equation}{0}
\setcounter{figure}{0}
\setcounter{table}{0}
\setcounter{page}{1}
\makeatletter
\renewcommand{\theequation}{S\arabic{equation}}
\renewcommand{\thefigure}{S\arabic{figure}}
\renewcommand{\bibnumfmt}[1]{[S#1]}
\renewcommand{\citenumfont}[1]{S#1}

\section{I. MODAL FREQUENCIES IN THE DEGENERATE CAVITY LASER}
\begin{figure}[h]
    \includegraphics{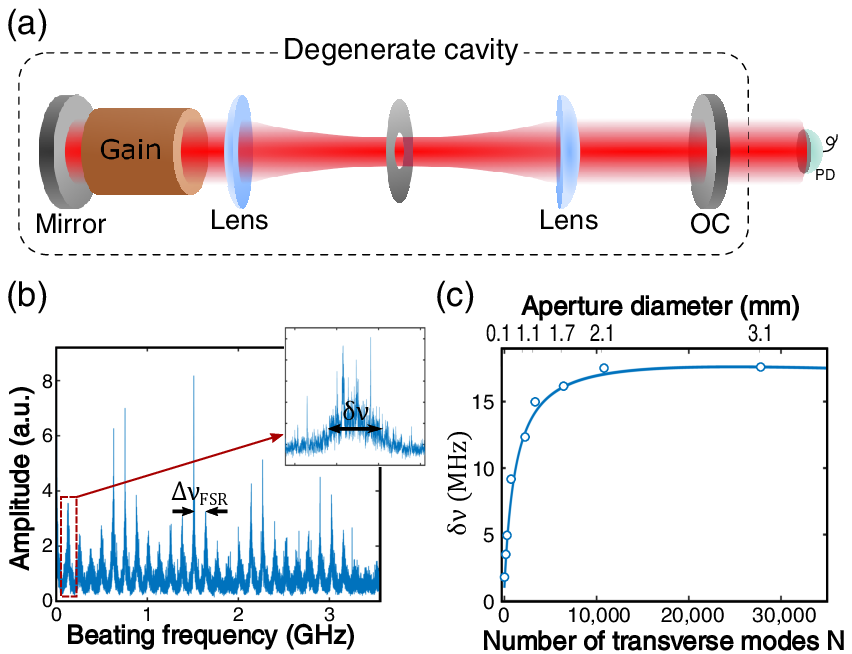}
	\caption{Experimental analysis of the modal frequency spectrum. (a) Experimental arrangement of the DCL, with a circular aperture at the Fourier plane, and a fast photodetector. (b) Beating frequency spectrum of the laser output, as detected by a fast photodetector. (c) Width of the first beating frequency as a function of the circular aperture diameter and the corresponding number of modes.}
	\label{fig:beating}
\end{figure}

The spectral width of all modes around a given longitudinal frequency in a non-ideal degenerate cavity laser (DCL) is expected to scale with the number of transverse modes supported by the cavity. The number of transverse modes can be varied by placing a variable circular aperture at the Fourier plane between the two lenses inside the DCL [1,2], as shown in Fig.~\ref{fig:beating}(a). The circular aperture serves as a spatial filter, introducing loss to higher order modes and their elimination due to mode competition. To verify the modal frequency spectrum in the manuscript, we first measured the intensity time dynamics of the laser output with a fast photodetector, for various aperture sizes. 

Taking a Fourier transform of the measured intensities, we found the beating frequency spectrum of the lasing modes, for various aperture sizes. Figure~\ref{fig:beating}(b) shows a representative beating frequency spectrum from the DCL. We then calculated the width of the beating frequency spectrum around the first beating frequency at $\Delta\nu_{FSR}=123\MHz$, for each aperture size. The width was determined by calculating the second moment of the beating frequency distribution. Figure~\ref{fig:beating}(c) shows the beating frequency spectral width $\delta\nu$ as a function of the number of lasing transverse modes $N$. 

Then, the number of modes was determined from the beam quality factor, which is the product of beam width at the output plane of the laser and the divergence angle of the beam, normalized by the product for a single mode laser output $\lambda/\pi$. As evident, for a small aperture and correspondingly a single transverse mode, the beating frequency spectral width is $\delta\nu=1.8\MHz$, which corresponds to a frequency width of $1.3\MHz$ (assuming Gaussian profiles). This value is expected to determine the point of transition between long and short exposure times, however it does not agree with our experimental data (see Fig.~\ref{fig:ExpData}(d) in the manuscript). Also evident, is a monotonic increase in the beating frequency spectral width with the number of lasing modes, in agreement with the predicted modal frequency spectrum, until it saturates for larger apertures and larger numbers of transverse modes.

\section{II. VALIDATION OF METHOD FOR SPECKLE CONTRAST MEASUREMENT}
The speckle contrast was determined in the manuscript from many uncorrelated time series measurements of a single pixel, with a fast photodetector. As noted there, each photodetector measurement was integrated in time, and the speckle contrast was calculated as $C(\Delta\tau)=[\langle\bar{I}(\Delta\tau)^2\rangle/\langle\bar{I}(\Delta\tau)\rangle^2-1]^{1/2}$, where $\langle\cdot\rangle$ denotes an ensemble average over different points in the speckle field and $\bar{I}(\Delta\tau)$ denotes the time average intensity, $\bar{I}(\Delta\tau)=\frac{1}{\Delta\tau}\int_0^{\Delta\tau} I(t)dt$. 

To verify the validity of such measurement method, we also performed speckle contrast measurements based on full 2D speckle images taken by a CMOS camera, in the limit of long exposure time ($\Delta\tau>50 \mu sec$). As in the previous section, a circular aperture was placed at the Fourier plane between the two lenses, and served to filter out higher order transverse modes. The results of speckle contrast as a function of aperture size are presented in Fig.~\ref{fig:aperture}. As evident, decreasing the aperture size, i.e. reducing the number of lasing transverse modes, increased the speckle contrast [1,2]. There is good agreement with the speckle contrast measurement method presented in the manuscript.

\begin{figure}[h]
    \includegraphics{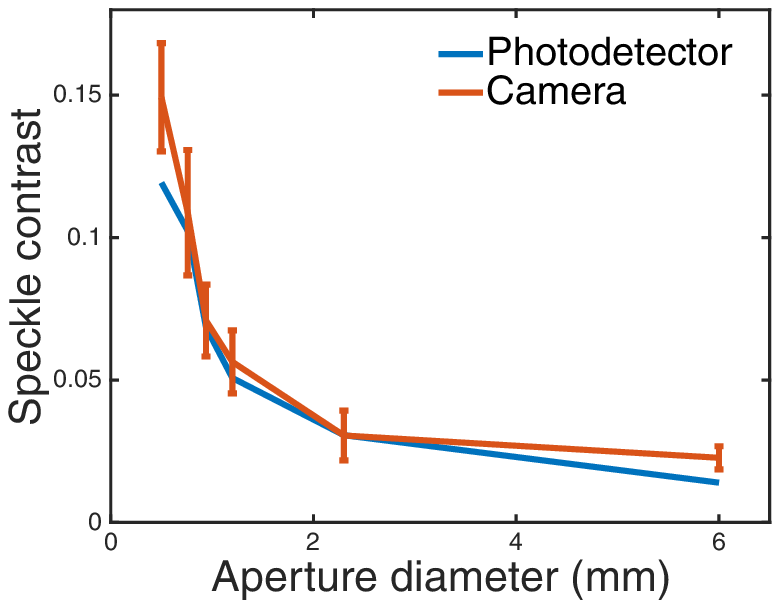}
	\caption{Speckle contrast as a function of circular aperture diameter, as calculated from a full 2D speckle image and as calculated from many single point sampling of the speckle field by a fast photodetector.}
	\label{fig:verify}
\end{figure} 

\section{III. EXPERIMENTAL SPECKLE CONTRAST MEASUREMENT IN THE REGIME $N\ll M$}
\begin{figure}[t]
    \includegraphics{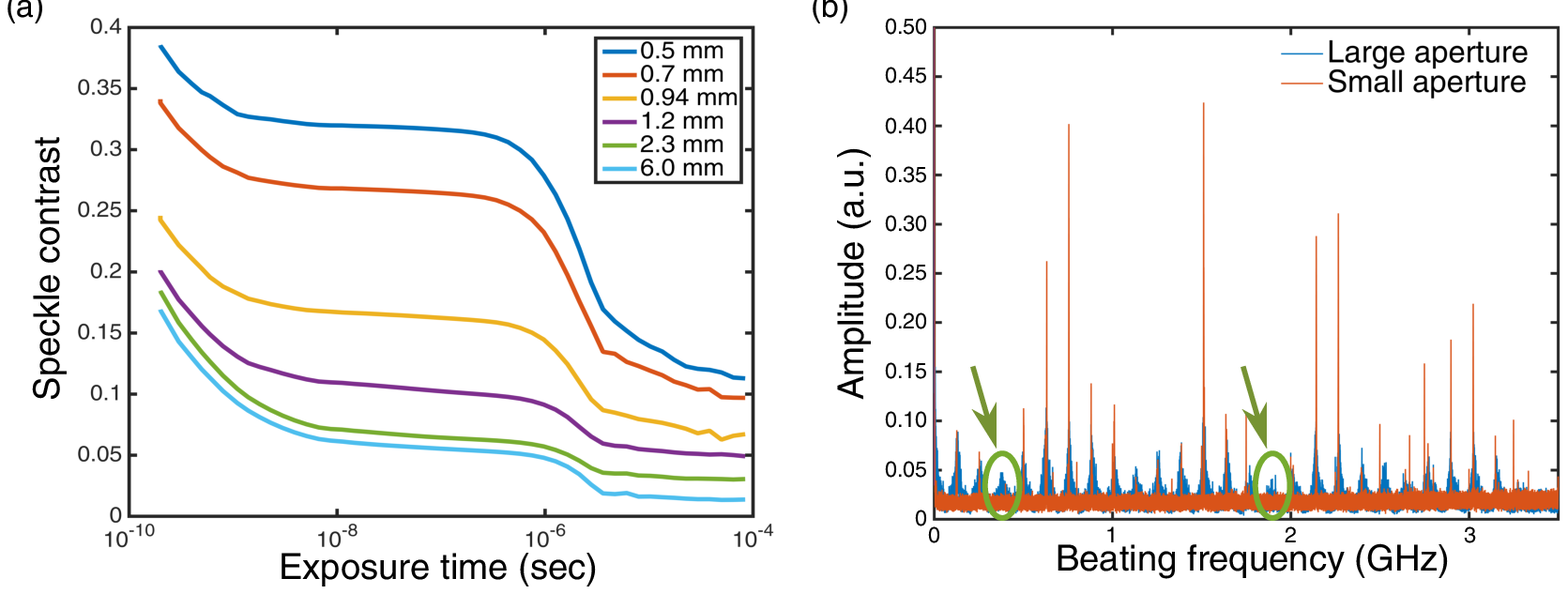}
	\caption{Effect of intra-cavity circular aperture size on speckle contrast time dynamics. (a) Speckle contrast as a function of exposure time for different diameters of the intra-cavity aperture. (b) Comparison of the beating frequency spectrum for small (0.22mm in diameter) and large (1.3mm in diameter) circular apertures.}
	\label{fig:aperture}
\end{figure}

As noted above, the number of transverse modes in the DCL can be adjusted by varying the size of a circular aperture placed at the Fourier plane between the two lenses. Figure~\ref{fig:beating}(a) shows speckle contrast as a function of exposure time for several different aperture diameters. In the limit of long exposure time $\Delta\tau$, the measured values of speckle contrast correspond well with the known relation $C=N^{-1/2}$, where the number of modes $N$ was determined by measuring the beam quality factor [1], as shown in section I. As evident, all graphs are similar to one another, and qualitatively agree with the expected behavior for $N\gg M$ (see Figs.~\ref{fig:ExpData}(d) and~\ref{fig:SimSpeckle}(d) in the manuscript). This is despite the great decrease in the number of transverse modes. Ideally, varying the size of the aperture should only affect the number of lasing transverse modes, and not the number of lasing longitudinal modes. Therefore, for small aperture sizes we expect the DCL to reach the regime of $N\ll M$. However, our experimental results indicate that this is not the case (compare Fig.~\ref{fig:aperture} to Fig.~\ref{fig:SimSpeckle}(e) in the manuscript). 

We attribute this behavior to the fact that the number of longitudinal modes for small apertures is lower than expected. To verify this, we measured the intensity time dynamics at the output of the DCL, using a fast photodetector. The resulting beating frequency spectra for two different sizes of apertures are presented in Fig.~\ref{fig:aperture}(b). As evident, several beating frequencies are missing in the case of a small aperture, indicating a significant difference in the number of lasing longitudinal modes. Since the results show the \textit{beating} frequencies, every peak is generated by a superposition of many longitudinal modes. A missing peak indicates that there are many longitudinal modes that are not lasing. For example, the third peak that is missing from the beating modal frequency spectrum for the case of a small aperture, indicates that at least a third of the longitudinal modes are missing. We therefore conclude that variations in the aperture size affect also the number of longitudinal modes in the laser cavity, and that it is likely that for all measurements in Fig.~\ref{fig:aperture}(a) $N\gg M$ . 

\section{IV. EXPECTED SPECKLE CONTRAST IN INTERMEDIATE SHORT EXPOSURE TIME}
\begin{figure}[h]
    \includegraphics{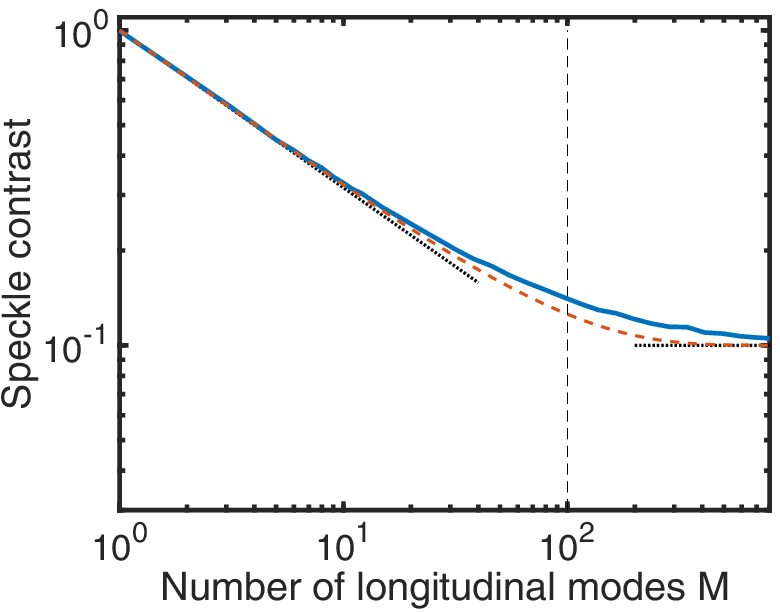}
	\caption{Speckle contrast at short exposure time as a function of the number of longitudinal modes in the DCL, for $N=100$ spatial modes. Numerical results (blue solid line) are in agreement with the analytic result (dashed red line). Vertical dashed line denotes the transition from  $M<N$ to $M>N$, and dotted black line denotes the expected asymptotic behavior for $M\ll N$ and $M\gg N$.}
	\label{fig:CVsM3}
\end{figure}

In this section we derive Eq.~\ref{eq:nu_qmn2} in the manuscript, determining the speckle contrast in the short exposure time regime. As noted above, in this regime the DCL has $M$ spatio-temporal supermodes, each of which is composed of $N$ spatial modes with random phases,

\begin{equation}
\label{eq:E_SpatialMode}
    E_{SM}^{(k)}=\sum_{j=1}^N E_j(x)e^{i\phi_j},
\end{equation}
where $E_{SM}^{(k)}$  is the $k^{th}$ supermode at location $x$, $E_j$ is the $j^{th}$ spatial mode at $x$, $\phi_j$ is the phase of the $j^{th}$ spatial mode and summation is taken over all $N$ spatial modes. 
For $M$ uncorrelated modes, the speckle contrast decays as $C=M^{-1/2}$. However, since the supermodes are statistically correlated, we must compute the average correlations between supermodes, and consider only the residues. The field correlation $\mu$ between two supermodes is
\begin{equation}
\label{eq:corr_mu}
    \mu=\frac{\int E_{SM}^{(1)}(x)E_{SM}^{*(2)}(x)dx}{\sqrt{\int I_{SM}^{(1)}(x)dx \int I_{SM}^{(2)}(x)dx}} = \frac{1}{\sqrt{N}} 
\end{equation}
and therefore the intensity correlation between two supermodes is [3]
\begin{equation}
    \rho=|\mu|^2=\frac{1}{N}.
\end{equation}
For $M$ supermodes, high order correlations must be taken into account, and the effective number of modes is 
\begin{equation}
    M_{eff} = \sum_{k=1}^M\binom{M}{k}\left(\frac{1}{N}\right)^{k-1}(-1)^{k+1}.
\end{equation}
Using the binomial theorem we find 
\begin{equation}
    M_{eff}=N\left[1-\left(1-\frac{1}{N}\right)^M\right],
\end{equation}
so the speckle contrast is
\begin{equation}
\label{eq:Meff}
    C=\frac{1}{\sqrt{N\left[1-\left(1-1/N\right)^M\right]}}.
\end{equation}
As noted in the manuscript, in the limit $M\ll N$, Eq.~\ref{eq:Meff} is approximated by $C=M^{-1/2}$, whereas for $M\gg N$ it is approximated by $C=N^{-1/2}$, corresponding to the classical limit.
Figure~\ref{fig:CVsM3} compares numerical and analytic results for speckle contrast in the short exposure time as a function of the number of longitudinal modes $M$, for $N=100$ spatial modes. The discrepancies are attributed to numerical errors. 

\section{Reference}
\noindent [1] M. Nixon, B. Redding,  a a Friesem, H. Cao, and N. Davidson, "Efficient method for controlling the spatial coherence of a laser," Opt. Lett. \textbf{38}, 3858–61 (2013).

\noindent [2] R. Chriki, M. Nixon, V. Pal, C. Tradonsky, G. Barach, A. A. Friesem, and N. Davidson, "Manipulating the spatial coherence of a laser source," Opt. Express \textbf{23}, 12989 (2015).

\noindent [3] J. W. Goodman, Speckle Phenomena in Optics: Theory and Applications (2005).

\end{document}